\documentclass[sigconf]{acmart}

\usepackage{booktabs}
\usepackage{balance}

\copyrightyear{2026}
\acmYear{2026}
\setcopyright{cc}
\setcctype{by}
\acmConference[CSCW Companion '26]{Companion of the Computer-Supported Cooperative Work and Social Computing}{October 10--14, 2026}{Salt Lake City, UT, USA}
\acmBooktitle{Companion of the Computer-Supported Cooperative Work and Social Computing (CSCW Companion '26), October 10--14, 2026, Salt Lake City, UT, USA}
\acmDOI{10.1145/3785651.3830025}
\acmISBN{979-8-4007-2378-0/2026/10}

\begin{document}

\title{A Social Norms Approach to Youth Social Media Design}

\author{JaeWon Kim}
\affiliation{%
  \institution{The Information School, University of Washington}
  \city{Seattle}
  \state{Washington}
  \country{USA}}
\email{jaewonk@uw.edu}

\begin{abstract}
Young people consistently say they want authentic self-expression, less judgment, and more interpersonal trust on social media, yet they rarely manage to engage that way. My dissertation argues that the obstacle is normative rather than individual: how youth engage is governed less by personal choice than by platform norms, peer perception, and beliefs about how others behave. I take a social norms approach to youth social media design organized around three claims. First, platform norms constrain individual behavior, producing a pluralistic ignorance in which youth enact norms they privately reject. Second, design interventions are themselves shaped by existing norms, so whether a feature works depends on the environment around it, which means relational goals such as privacy must be treated as social norms rather than individual settings. Third, a societal norm about what ``social media'' is---equating it with a few mainstream platforms---confines policy and design to mitigating those platforms rather than actively envisioning supportive alternatives. Together these claims motivate my dissertation research: engaging youth directly in designing and building an evidence-based independent platform whose features consistently signal that building trusted connections is what the space is for.
\end{abstract}



\maketitle

\section{Introduction and Motivation}
Young people are clear about what they want from social media and largely unable to get it. Across studies, adolescents and young adults say they want authentic self-expression, everyday sharing with people they trust, and connection without constant evaluation~\cite{kim2025trust, davis2025, landesman2024}. They can also name what gets in the way, from follower counts that invite comparison to the sense that posting anything ordinary looks like a bid for attention. What they cannot easily do is act on this knowledge, because how one engages online does not feel like a personal choice. It feels like a response to what the platform rewards, how peers might react, and how everyone else appears to be behaving~\cite{kim2025privacy, kim2024bereal}.

My dissertation takes this gap as its central problem and argues that it is normative. Youth often cannot engage as they want not because they lack motivation or awareness, but because the normative environment, shaped by platform design, discourages the behaviors they value. Changing what young people do online therefore means changing the norms that govern their behavior, not only the features they are offered. I pursue this through a social norms approach built on three claims, moving from how norms constrain individuals, to how they shape design, to how a norm about what social media is forecloses alternatives. Taken together, these point toward a platform built with youth whose design consistently communicates that trusted connection is what the space is for, and the sections below develop that argument and the research it motivates.

\section{Theoretical Grounding}
Social norms are informal understandings about what people do and what they ought to do, and they shape behavior even when they diverge from private preference. Bicchieri treats norms as conditional preferences sustained by two beliefs: empirical expectations about what others do, and normative expectations about what others think one should do~\cite{Bicchieri-2017-NormsWildNorms-b}. This structure explains why people conform to norms they privately reject, and why such norms can collapse once the underlying beliefs shift. Cialdini's focus theory separates descriptive norms, what others do, from injunctive norms, what others approve of, and holds that norms guide behavior mainly when salient~\cite{Cialdini-1990-FocusTheoryNormative-a}---a distinction that maps directly onto platform features, since metrics foreground what people do while platform defaults signal what they should do. And because norms are learned by observing others, with visible and higher-status behavior weighted most heavily~\cite{Bandura2001-xu}, whose behavior a platform makes visible matters as much as which behavior it makes visible.

This work also makes norms a practical lever for change. The social norms approach shifts behavior by correcting misperceptions and amplifying constructive norms rather than persuading individuals one at a time, reducing college drinking~\cite{Berkowitz-2005-OverviewSocialApproach-a} and peer conflict in schools~\cite{Paluck-2016-ChangingClimatesSchools-o}, and a consistent lesson is that changing perceived norms can precede behavioral change without first changing private attitudes~\cite{Tankard-2016-NormPerceptionChange-p}. The levers of this approach---what behavior is visible, what defaults apply, how actions are framed---are all design choices. Computer-mediated communication (CMC) theory explains why such choices carry weight: mediated features do not merely transmit social information but transform how interaction unfolds~\cite{walther1996}, and affordances enable and constrain action without determining it~\cite{evans2017}. Read alongside the norms literature, these traditions support the premise of my dissertation, that platform design does not simply host youth interaction but constitutes the normative environment in which it occurs.

\section{Three Ways Social Norms Shape Youth Social Media Design}
These claims build on one another. Because platform norms constrain individual behavior, interventions must target the normative environment rather than the individual user. Because a feature's effect depends on the norms around it, changing that environment takes a platform whose features cohere, not better features added one at a time. And as long as ``social media'' means a handful of mainstream apps, such alternatives are rarely imagined, let alone built. I develop each claim below, grounding it in completed and ongoing studies and drawing out the research it motivates.

\subsection{Platform Norms Constrain Individual Behavior}
Teens often know how they want to engage online and hold back anyway. An earnest post might look like trying too hard, a boundary might come across as uptight, a string of photos from a good weekend might read as bragging---so the safe move is whatever seems least likely to draw judgment. This is pluralistic ignorance: individuals privately reject a norm while assuming their peers endorse it, so everyone enacts a norm that few actually embrace~\cite{Bicchieri-2017-NormsWildNorms-b, Tankard-2016-NormPerceptionChange-p}.

This pattern appears in how teens manage privacy, decide what to disclose, and present themselves online. Among teens with public accounts, $28\%$ reported privacy fear that diminished their quality of life, yet they could not act on their own preferences without a supportive environment, and peers' expectations often pushed them toward conformity rather than protection~\cite{kim2025privacy}. Withdrawal is another common response: when an audience's trustworthiness is ambiguous, teens default to disengagement, not because they prefer silence but because the perceived social risk of disclosure is too high~\cite{kim2025trust}. By contrast, teens valued how BeReal nudged everyone toward casual, uncurated sharing, a change they wanted but could not produce individually on platforms built around performance~\cite{kim2024bereal}.

Together, these findings suggest the obstacle is normative rather than individual: what youth do follows from what they believe others do and approve of, and that belief is often mistaken. If individual choice is not the lever for change, design that reshapes the environment is.

\subsection{Design Interventions Are Shaped by Existing Norms}
No feature carries its intended effect on its own; what it accomplishes depends on the norms already in place. A privacy control that performs well in isolation can fail on an account where the environment feels vaguely hostile and a persistent, low-grade fear lingers regardless of the settings applied~\cite{kim2025privacy}. For a design to support a behavior, the environment around it must support that behavior too, which is why privacy is better treated as a social norm the whole platform upholds than as a setting each user manages alone~\cite{kim2025privacy}.

The same holds for trust: platform norms shape whether disclosure feels safe~\cite{kim2025trust}. BeReal's features pointed consistently toward ``being real''---synchronized prompts, ephemerality, reciprocal posting---so users absorbed that norm even when they joined without it, shared in ways they would not elsewhere, and felt able to hold one another to it~\cite{kim2024bereal}. When other platforms later adopted similar features without an equally clear signal (TikTok Now~\cite{tiktok2022}), the same designs worked less well; a feature borrowed out of its normative context does not reproduce its effect. A design supports a behavior, then, when its features pull in one direction and make the platform's purpose unmistakable.

This also changes where design effort should go. If features only work when the surrounding norms support them, adding better features to mainstream platforms that youth already experience as toxic is unlikely to succeed; the field needs evidence-based alternative platforms instead. The platforms youth themselves describe as friendship-supportive, such as Discord, stand out not for any single feature but for the overall environment---persistent spaces, communities that set their own rules, and trust both among users and between users and the people who run the platform~\cite{kim2025discord}. People seeking different norms may need different platforms.

\subsection{Fixation on Mainstream Platforms Forecloses Alternatives}
The third norm operates not in any specific platform's design but in public discourse: ``social media'' has come to mean a handful of mainstream platforms. Because Instagram and TikTok essentially stand in for social media as a whole, their feed-driven, engagement-optimized paradigm passes for what social media simply is, and policy responds with prevention, restriction, and bans aimed at those platforms~\cite{kim2026hogwarts, weinstein2022}. Because the paradigm is also considered harmful, young people's heavy use of it invites a familiar inference: if social media is bad and youth use it anyway, then youth must not know what is good for them. Youth report that this framing leaves them feeling unheard, yet they often absorb it too, describing social media as bad and asking for less doomscrolling while their own behavior tells a more complicated story~\cite{kim2026hogwarts}. The framing thus forecloses the more constructive and realistic question of what else social media could be.

Treating the mainstream platforms' design patterns as fixed also obscures what digital spaces now provide. With physical third places increasingly scarce, online environments are where much of young people's social life happens, so removing access removes connection without restoring anything healthier~\cite{kim2025discord}. The loss lands hardest on younger adolescents, whose offline world may extend no further than a single school, and on youth with niche identities, interests, or disabilities, who may find few compatible peers nearby; research on queer joy, hospitalized children, and niche-interest communities shows how much these groups gain from connecting across distance~\cite{kim2025discord, steeds2025, freeman2022, schaadhardt2023, lee2022, fiesler2017GrowingTheir, to2023, ellison2007}. Technology that links people across time and space has real potential when designed well, and ``social media'' need not mean the platforms that currently monopolize the term.

Focusing only on preventing harm misses technology's potential to support flourishing, since the absence of harm is not the presence of good~\cite{kim2024positech, kim2025positech}. Once the vocabulary of existing platforms is set aside, that potential becomes visible: asked how wizards might stay close across distance rather than how to fix social media, youth move from reactive problem-fixing to visions of ambient, playful, low-stakes connection and describe the exercise as hopeful~\cite{kim2026hogwarts}. That is the direction my research takes: engaging youth to imagine and build alternatives rather than only mitigating the platforms that already exist.

\section{Work to Date and Next Steps}
My dissertation develops these claims through qualitative, design, and field studies with youth ages 13--25~\cite{kim2026attunement}. Completed work grounds all three~\cite{kim2025privacy, kim2024bereal, kim2025trust, kim2025discord, kim2026hogwarts}. The ongoing deployments are a product of these findings and a test of them. I built a friendship-supportive platform and ran a four-week within-subjects crossover deployment with $99$ youth ages 15--25 across Korea and the United States, alternating between a control version resembling mainstream platforms and an experimental version whose features were derived from the trust-enabled privacy framework~\cite{kim2025trust, kim2026deployment}. Combining behavioral logs with interviews, this work specifies the design mechanisms through which norms were signaled, transferred, or disrupted. An ongoing follow-up deployment responds to a clear lesson from the first: too many options made it hard for users to tell what was normative (what the platform intended and what they are expected to do), so the follow-up sharpened those signals to test whether clearer norms yield more consistent behavior.

Planned work carries the third claim forward. I am developing an Asynchronous Remote Community study in which youth collectively articulate goals, criteria, and boundary conditions for a healthier social media environment, defining for themselves what ``better'' should mean rather than reacting to harms defined for them. The study is also designed to build participants' understanding of how design shapes norms, which norms support which kinds of interaction, and what to expect from a given platform, so they make more informed choices and know more clearly what they want when seeking alternatives.

\section{Anticipated Contributions}
This dissertation makes three contributions to CSCW. First, it applies the social norms approach to mediated settings, giving an empirically grounded account of how platform design creates and sustains the normative environments that govern youth behavior, where prior norms research has largely theorized offline contexts. Second, it shows why evaluating design interventions feature by feature in isolation is less effective, and reframes relational goals such as privacy and trust as social norms an entire environment must consistently signal and support, mapping the design spaces through which platforms can do so~\cite{kim2026designspace}. Third, it makes the case for building developmentally supportive alternatives with youth rather than just mitigating mainstream platforms, expanding the problem space of youth social media by offering a generative rather than purely protective path forward~\cite{weinstein2022, kim2025positech}.

\begin{acks}
I thank my advisor, Alexis Hiniker, and my committee members Katie Davis, Casey Fiesler, Amy X. Zhang, and Sean Munson for their guidance. I also acknowledge the CERES Network, University of Washington Global Innovation Funds (GIF), and Student Technology Funds (STF), which supported this work, along with the Paul G. Allen School of Computer Science \& Engineering Endowed Fund for Excellence and a gift from Google.
\end{acks}

\balance

\bibliographystyle{ACM-Reference-Format}
\bibliography{references, newrefs, references1, references2, references3, refs}

@inproceedings{kim2025positech,
author = {Kim, JaeWon and Liu, Jiaying and Popowski, Lindsay and Pyle, Cassidy and Arif, Ahmer and Hayes, Gillian R. and Hiniker, Alexis and Ju, Wendy and Mueller, Florian Floyd and Shen, Hua and Somanath, Sowmya and Fiesler, Casey and Kotturi, Yasmine},
title = {Design for Hope: Cultivating Deliberate Hope in the Face of Complex Societal Challenges},
year = {2025},
isbn = {9798400714801},
publisher = {Association for Computing Machinery},
address = {New York, NY, USA},
url = {https://doi.org/10.1145/3715070.3748287},
doi = {10.1145/3715070.3748287},
booktitle = {Companion Publication of the 2025 Conference on Computer-Supported Cooperative Work and Social Computing},
pages = {112–115},
numpages = {4},
location = {
},
series = {CSCW Companion '25}
}

@article{kim2025privacy,
author = {Kim, JaeWon and Cho, Soobin and Wolfe, Robert and Nair, Jishnu Hari and Hiniker, Alexis},
title = {Privacy as Social Norm: Systematically Reducing Dysfunctional Privacy Concerns on Social Media},
year = {2025},
issue_date = {May 2025},
publisher = {Association for Computing Machinery},
address = {New York, NY, USA},
volume = {9},
number = {2},
url = {https://doi.org/10.1145/3711049},
doi = {10.1145/3711049},
journal = {Proc. ACM Hum.-Comput. Interact.},
month = may,
articleno = {CSCW151},
numpages = {39}
}

@article{kim2024bereal,
author = {Kim, JaeWon and Wolfe, Robert and Chordia, Ishita and Davis, Katie and Hiniker, Alexis},
title = {"Sharing, Not Showing Off": How BeReal Approaches Authentic Self-Presentation on Social Media Through Its Design},
year = {2024},
issue_date = {November 2024},
publisher = {Association for Computing Machinery},
address = {New York, NY, USA},
volume = {8},
number = {CSCW2},
url = {https://doi.org/10.1145/3686909},
doi = {10.1145/3686909},
journal = {Proc. ACM Hum.-Comput. Interact.},
month = nov,
articleno = {370},
numpages = {32}
}

@inproceedings{kim2024positech,
author = {Kim, JaeWon and Popowski, Lindsay and Fang, Anna and Pyle, Cassidy and Freeman, Guo and Kelly, Ryan M. and Lee, Angela Y. and Liu, Fannie and Smith, Angela D. R. and To, Alexandra and Zhang, Amy X.},
title = {Envisioning New Futures of Positive Social Technology: Beyond Paradigms of Fixing, Protecting, and Preventing},
year = {2024},
isbn = {9798400711145},
publisher = {Association for Computing Machinery},
address = {New York, NY, USA},
url = {https://doi.org/10.1145/3678884.3681833},
doi = {10.1145/3678884.3681833},
booktitle = {Companion Publication of the 2024 Conference on Computer-Supported Cooperative Work and Social Computing},
pages = {701–704},
numpages = {4},
location = {San Jose, Costa Rica},
series = {CSCW Companion '24}
}

@misc{kim2026hogwarts,
      title={Social Media Should Feel Like Minecraft, Not Instagram: Youth Visions for Meaningful Social Connections through Fictional Inquiry}, 
      author={JaeWon Kim and Hyunsung Cho and Fannie Liu and Alexis Hiniker},
      year={2026},
      eprint={2502.06696},
      archivePrefix={arXiv},
      primaryClass={cs.HC},
      url={https://arxiv.org/abs/2502.06696}, 
}

@inproceedings{kim2025trust,
author = {Kim, JaeWon and Wolfe, Robert and Subramanian, Ramya Bhagirathi and Lee, Mei-Hsuan and Colnago, Jessica and Hiniker, Alexis},
title = {Trust-enabled privacy: social media designs to support adolescent user boundary regulation},
year = {2025},
isbn = {978-1-939133-51-9},
publisher = {USENIX Association},
address = {USA},
booktitle = {Proceedings of the Twenty-First USENIX Conference on Usable Privacy and Security},
articleno = {27},
numpages = {20},
location = {Seattle, WA, USA},
series = {SOUPS '25}
}

@misc{kim2025discord,
      title={Discord's Design Encourages "Third Place" Social Media Experiences}, 
      author={JaeWon Kim and Thea Klein-Balajee and Ryan M. Kelly and Alexis Hiniker},
      year={2025},
      eprint={2501.09951},
      archivePrefix={arXiv},
      primaryClass={cs.HC},
      url={https://arxiv.org/abs/2501.09951}, 
}

@inproceedings{kim2026attunement,
  title        = {Designing Youth Social Media through Problem Space Attunement},
  author       = {Kim, JaeWon},
  year         = 2026,
  booktitle    = {Designing Interactive Systems Conference (DIS Companion '26)},
  publisher    = {Association for Computing Machinery},
  address      = {New York, NY, USA},
  doi          = {10.1145/3802974.3807979}
}

@inproceedings{kim2026designspace,
  title        = {Social Understanding, Placeness, and Identity Alignment: A Design Framework for Friendship-Supportive Youth Social Media},
  author       = {Kim, JaeWon and Hiniker, Alexis},
  year         = 2026,
  booktitle    = {Proceedings of the 25th Interaction Design and Children Conference (IDC '26)},
  publisher    = {Association for Computing Machinery},
  address      = {New York, NY, USA},
  doi          = {10.1145/3773077.3812190}
}

@inproceedings{kim2026deployment,
author = {Kim, JaeWon},
title = {Mapping the Design Space for Youth Social Media: A Framework Centered on Friendship Building},
year = {2026},
isbn = {9798400722837},
publisher = {Association for Computing Machinery},
address = {New York, NY, USA},
url = {https://doi.org/10.1145/3773077.3813777},
doi = {10.1145/3773077.3813777},
booktitle = {Proceedings of the 25th Annual ACM Interaction Design and Children Conference},
pages = {1406–1409},
numpages = {4},
location = {
},
series = {IDC '26}
}

@article{ellison2007,
  _source_file  = {my_refs/references (10).bib},
  author        = {Ellison, Nicole B and Steinfield, Charles and Lampe, Cliff},
  doi           = {10.1111/j.1083-6101.2007.00367.x},
  issn          = {1083-6101},
  journal       = {Journal of computer-mediated communication},
  language      = {en},
  month         = {July},
  number        = {4},
  pages         = {1143--1168},
  publisher     = {Oxford University Press (OUP)},
  title         = {The benefits of Facebook ``friends:'' Social capital and college students' use of online social network sites},
  volume        = {12},
  year          = {2007}
}

@article{evans2017,
  _source_file  = {my_refs/references (10).bib},
  abstract      = {This study aims to clarify inconsistencies regarding the term affordances by examining how affordances terminology is used in empirical research on communication and technology. Through an analysis of 82 communication-oriented scholarly works on affordances, we identify 3 inconsistencies regarding the use of this term. First, much research describes a particular affordance without engaging other scholarship addressing that affordance. Second, several studies identify ``lists'' of affordances without conceptually developing individual affordances within those lists. Third, the affordances perspective is evoked in situations where the purported affordance does not meet commonly accepted definitions. We conclude with a set of criteria to aid scholars in evaluating their assumptions about affordances and to facilitate a more consistent approach to its conceptualization and application.},
  author        = {Evans, Sandra K and Pearce, Katy E and Vitak, Jessica and Treem,
Jeffrey W},
  doi           = {10.1111/jcc4.12180},
  issn          = {1083-6101},
  journal       = {J Comput-Mediat Comm},
  language      = {en},
  number        = {1},
  pages         = {35--52},
  title         = {Explicating Affordances: A Conceptual Framework for Understanding
Affordances in Communication Research: {EXPLICATING} {AFFORDANCES}},
  volume        = {22},
  year          = {2017}
}

@inproceedings{schaadhardt2023,
  _source_file  = {my_refs/references (10).bib},
  abstract      = {Today's youth face many mental health challenges and are increasingly represented in psychiatric hospitalizations. Scholars have sought to understand social media's role in mental health issues, but limited work has explored TikTok---the video-centric social media platform that is popular with youth---and people's connections around psychiatric hospitalization experiences. In this study, we used qualitative content analysis to examine a random sample of 140 TikTok posts related to psychiatric hospitalization. We found that members of this population frequently utilize humor to create and maintain a positive and supportive community with each other. We also describe how TikTok's design affords these interactions among community members, and conclude with a series of provocations for researchers and designers working at the intersections of social media and mental illness. We hope our study provides insights for how to further support rather than just censor youth in using creative outlets to connect with each other.},
  address       = {New York, NY, USA},
  author        = {Schaadhardt, Anastasia and Fu, Yue and Pratt, Cory Gennari and Pratt, Wanda},
  booktitle     = {Proceedings of the 2023 CHI Conference on Human Factors in Computing Systems},
  doi           = {10.1145/3544548.3581559},
  language      = {en},
  month         = {April},
  pages         = {1--13},
  publisher     = {ACM},
  title         = {``Laughing so {I} don't cry'': How {TikTok} users employ humor and compassion to connect around psychiatric hospitalization},
  volume        = {14},
  year          = {2023}
}

@article{walther1996,
  _source_file  = {my_refs/references (10).bib},
  abstract      = {While computer-mediated communication use and research are proliferating rapidly, findings offer contrasting images regarding the interpersonal character of this technology. Research trends over the history of these media are reviewed with observations across trends suggested so as to provide integrative principles with which to apply media to different circumstances. First, the notion that the media reduce personal influences?their impersonal effects?is reviewed. Newer theories and research are noted explaining normative ?interpersonal? uses of the media. From this vantage point, recognizing that impersonal communication is sometimes advantageous, strategies for the intentional depersonalization of media use are inferred, with implications for Group Decision Support Systems effects. Additionally, recognizing that media sometimes facilitate communication that surpasses normal interpersonal levels, a new perspective on ?hyperpersonal? communication is introduced. Subprocesses are discussed pertaining to receivers, senders, channels, and feedback elements in computer-mediated communication that may enhance impressions and interpersonal relations.},
  author        = {Walther, Joseph B.},
  doi           = {10.1177/009365096023001001},
  file          = {SAGE PDF Full Text:/Users/jaewonk/Zotero/storage/TJPBMEHU/WALTHER - 1996 - Computer-Mediated Communication Impersonal, Inter.pdf:application/pdf},
  issn          = {0093-6502},
  journal       = {Communication Research},
  language      = {en},
  month         = {February},
  note          = {Publisher: SAGE Publications Inc},
  number        = {1},
  pages         = {3--43},
  publisher     = {SAGE Publications Inc},
  shorttitle    = {Computer-{Mediated} {Communication}},
  title         = {Computer-{Mediated} {Communication}: {Impersonal}, {Interpersonal}, and {Hyperpersonal} {Interaction}},
  url           = {https://doi.org/10.1177/009365096023001001},
  urldate       = {2022-12-07},
  volume        = {23},
  year          = {1996}
}

@book{weinstein2022,
  _source_file  = {my_refs/references (10).bib},
  abstract      = {How teens navigate a networked world and how adults can support them.What are teens actually doing on their smartphones? Contrary to many adults' assumptions, they are not simply ``addicted'' to their screens, oblivious to the afterlife of what they post, or missing out on personal connection. They are just trying to navigate a networked world. In Behind Their Screens, Emily Weinstein and Carrie James, Harvard researchers who are experts on teens and technology, explore the complexities that teens face in their digital lives, and suggest that many adult efforts to help---``Get off your phone!'' ``Just don't sext!''---fall short. Weinstein and James warn against a single-minded focus by adults on ``screen time.'' Teens worry about dependence on their devices, but disconnecting means being out of the loop socially, with absence perceived as rudeness or even a failure to be there for a struggling friend. Drawing on a multiyear project that surveyed more than 3,500 teens, the authors explain that young people need empathy, not exasperated eye-rolling. Adults should understand the complicated nature of teens' online life rather than issue commands, and they should normalize---let teens know that their challenges are shared by others---without minimizing or dismissing. Along the way, Weinstein and James describe different kinds of sexting and explain such phenomena as watermarking nudes, comparison quicksand, digital pacifiers, and collecting receipts. Behind Their Screens offers essential reading for any adult who cares about supporting teens in an online world.},
  author        = {Weinstein, Emily and James, Carrie},
  isbn          = {9780262047357},
  language      = {en},
  month         = {August},
  publisher     = {MIT Press},
  title         = {Behind Their Screens: What Teens Are Facing (and Adults Are
Missing)},
  year          = {2022}
}

@inproceedings{davis2025,
  _source_file  = {my_refs/references (9).bib},
  abstract      = {Prior work has documented various ways that teens use social
media to regulate their emotions. However, little is known about
what these processes look like on a moment-by-moment basis. We
conducted a diary study to investigate how teens (N= 57, Mage=
16.3 years) used Instagram to regulate their emotions. We
identified three kinds of emotionally-salient drivers that
brought teens to Instagram and two types of behaviors that
impacted their emotional experiences on the platform. Teens
described going to Instagram to escape, to engage, and to manage
the demands of the platform. Once on Instagram, their primary
behaviors consisted of mindless diversions and deliberate acts.
Although teens reported many positive emotional responses, the
variety, unpredictability, and habitual nature of their
experiences revealed Instagram to be an unreliable tool for
emotion regulation (ER). We present a model of teens’ ER
processes on Instagram and offer design considerations for
supporting adolescent emotion regulation.},
  author        = {Davis, Katie and Landesman, Rotem and Yoon, Jina and Kim, JaeWon
and Muñoz Lopez, Daniela E and Magis-Weinberg, Lucía and Hiniker,
Alexis},
  booktitle     = {ACM CHI Conference on Human Factors in Computing Systems (CHI
’25)},
  publisher     = {ACM},
  title         = {“You Go Through So Many Emotions Scrolling Through Instagram”:
How Teens Use Instagram To Regulate Their Emotions},
  year          = {2025}
}

@article{freeman2022,
  _source_file  = {my_refs/references (1).bib},
  abstract      = {This paper focuses on embodied visibility emerging in social
Virtual Reality (VR) as a new lens to explore how queer users
build and experience visibility in nuanced ways. Drawing on 29
queer social VR users' experiences across various countries and
cultures, we identify three main strategies for building and
experiencing embodied visibility in social VR, limitations of
each strategy, and impacts of such visibility on queer users'
identity practices online and offline. We broaden current studies
on queer visibility online and expand the traditional lens of
selective visibility by highlighting how embodiment both supports
and challenges the multidimensional online presentations of queer
identity. We also propose potential design considerations to
further support diverse queer users' visibility in social VR and
inform future directions for creating inclusive online social
experiences.},
  address       = {New York, NY, USA},
  author        = {Freeman, Guo and Acena, Dane},
  doi           = {10.1145/3555153},
  journal       = {Proc. ACM Hum.-Comput. Interact.},
  keywords      = {virtual reality, social, selective visibility, queer identity,
online visibility, embodiment},
  month         = {November},
  number        = {CSCW2},
  pages         = {1--32},
  publisher     = {Association for Computing Machinery},
  title         = {``Acting Out'' Queer Identity: The Embodied Visibility in Social
Virtual Reality},
  volume        = {6},
  year          = {2022}
}

\end{document}